\documentclass[12pt,a4paper]{article}
\usepackage{graphicx,caption}
\usepackage[stable]{footmisc}
\usepackage{times}
\newcommand{\farcs}{\mbox{\ensuremath{.\!\!^{\prime\prime}}}}%

\title{\bf Massive stars in the era of ELTs}
\author{Christopher J. Evans$^1$\\
\vspace{1cm}\\
\normalsize $^1$ UK Astronomy Technology Centre, Blackford Hill, Edinburgh, EH9 3HJ, UK\\ }
\date{\mbox{}}
\begin{document}
\maketitle
\pagestyle{empty}
%
%
\def\bull{\vrule height .9ex width .8ex depth -.1ex}
\makeatletter
\def\ps@plain{\let\@mkboth\gobbletwo
\def\@oddhead{}\def\@oddfoot{\hfil\tiny\bull\quad
``The multi-wavelength view of hot, massive stars''; 39$^{\rm th}$ Li\`ege Int.\ Astroph.\ Coll., 12-16 July 2010 \quad\bull}%
\def\@evenhead{}\let\@evenfoot\@oddfoot}
\makeatother
%
%
\def\beginrefer{\section*{References}%
\begin{quotation}\mbox{}\par}
\def\refer#1\par{{\setlength{\parindent}{-\leftmargin}\indent#1\par}}
\def\endrefer{\end{quotation}}
%
%
{\noindent\small{\bf Abstract:} 
Plans for the next generation of optical-infrared telescopes, the
Extremely Large Telescopes (ELTs), are well advanced. With
primary apertures in excess of 20\,m, they will revolutionise our
ground-based capabilities. In this review I summarise the three current
ELT projects, their instrumentation plans, and discuss their science 
case and potential performance in the context of studies of massive
stars.}
%
%
\section{Introduction}
In many fields of astronomical research we have already approached the
sensitivity limits of the 8-10\,m class telescopes.  Improvements in
performance such as better instrument throughputs and further
developments in adaptive optics (AO) will likely come in the next few years,
but we are fundamentally limited by the collecting areas and
(potential) diffraction limits of existing facilities.

The primary motivation to embark on plans for the next generation of
ground-based optical and near-IR facilities, the Extremely Large
Telescopes (ELTs), is the huge gain from the combination of
sensitivity and angular resolution.  Large primary apertures will
collect more photons from each target and, with correction of
atmospheric turbulence via AO, we will achieve angular resolutions
beyond our current capabilties (excluding those targets which are
bright enough for optical-IR interferometry).  The science case for
the ELTs is typically split into three areas:
\begin{itemize}
\item{{\it Planets \& Stars:} including detection and characterization
of extra-solar planets, and a broad range of topics relating to solar system
science, proto-planetary discs and star formation.}
\item{{\it Star \& Galaxies:} bridging `local' topics such as the formation and evolution
of stellar clusters, out to using stellar populations to study the assembly histories of galaxies beyond the
Local Group and, to even larger distances, the study of black holes in active galactic nuclei.}
\item{{\it Galaxies \& Cosmology:} ranging from studies of galaxy evolution at low redshifts, out
to characterization of the highest redshift (`first light') systems and studies of cosmic expansion.}
\end{itemize}

Some of the cases relating to ELT studies of massive stars are now
discussed, together with a brief consideration of the broader
astronomical landscape in the 2020s.  Section~\ref{elt} gives an
overview of the three current ELT projects, while
Section~\ref{performances} gives example performances and highlights
the need to improve near-IR diagnostics to exploit the ELTs to their
best potential.  Lastly, Section~\ref{mad} summarises recent
observations with a successful AO pathfinder on the Very Large
Telescope (VLT), as an example of the power of `wide field' AO
techniques.

\subsection{Star formation}
The main thrust of this review is related to extra-galactic studies
but it is worth noting the likely contribution of the ELTs toward our
understanding of star formation.  The tremendous gain here will be the
combination of angular resolution with the ability to penetrate the
significant optical extinction (typically tens to over a hundred
magnitudes) toward ultra-compact H\,{\footnotesize II} regions by
observing at near- and mid-IR wavelengths (Zinnecker, 2006). Together
with other new facilities such as the Atacama Large Millimetre Array
(ALMA), we will truly have a multi-wavelength suite of tools to probe
the processes at work in the formation of massive stars.

\begin{center}
\begin{figure}[h]
\vspace{-0.25in}
\centering
\includegraphics{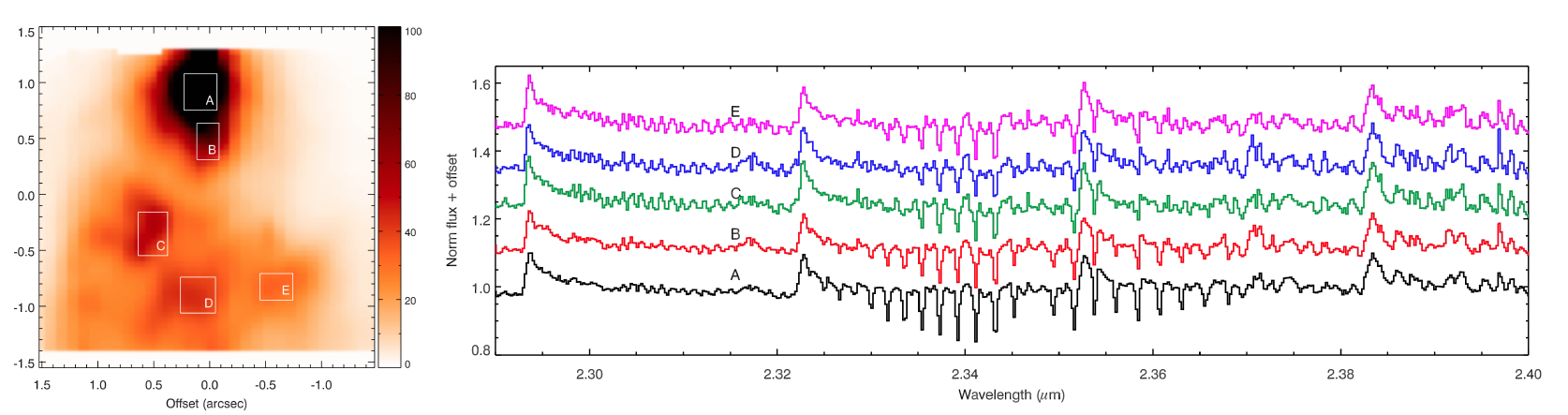}
\caption{{\it Left:} Gemini-NIFS $K$-band image of W33A from Davies et al. (2010).
{\it Right:} $K$-band spectra of the bright central
source (region `A') and other knots of emission (`B' to `E').  The variations in 
the CO spectra were used by Davies et al. to constrain the geometry of 
material around W33A.\label{nifs}}
\end{figure}
\end{center}

The potential of modern techniques in this area is neatly illustrated by recent
observations of W33A by Davies et al. (2010).  This is a well-studied
massive young stellar object in the Galaxy, at a distance of nearly
4\,kpc and with significant extinction at optical wavelengths.  Using
AO-corrected, $K$-band observations with the Gemini Near-IR Integral
Field Spectrograph (NIFS), Davies et al. were able to investigate the
different spatial structures related to W33A. Supported by
interferometric results from de Wit et al. (2007, 2010),
they conclude that the star is forming via similar processes to those
seen in lower mass objects, i.e., accretion from a circumstellar disc,
while driving a bi-polar outflow.  The collapsed $K$-band NIFS image
is shown in the left-hand panel of Fig.~\ref{nifs}, with spectra of
selected regions shown in the right-hand panel; note the spatial
scales of $\sim$0\farcs1 in the NIFS image.

With much greater sensitivity, the ELTs will be able to extend studies
of ultra-compact H\,{\footnotesize II} regions and young stellar
objects/clusters to much larger distances and to systems with greater
extinction.  In this area there will be strong complementarity between
the scales probed by the ELTs and by interferometry with, e.g, the
Very Large Telescope Interferometer (VLTI).  Indeed, the power of VLTI
is highlighted by the recent detection of a dusty disc around another
very young massive star, IRAS 13481$-$6124 (Kraus et al. 2010).

Although we are starting to glean some of the first views of the
formation of individual massive stars, the majority appear to be in
binary/multiple systems (e.g., Mason et al. 2009; Sana
\& Evans, 2010).  The dominant formation mechanism of such systems is still
not clear (e.g. Gies, 2008) and, if we want to be able to understand the integrated
light populations of distant star-forming galaxies (which are dominated by massive
stars), it is clear we still have much to learn.

\subsection{Stellar spectroscopy beyond the Local Group}

One of the big impacts of the 8-10\,m telescopes compared to previous
facilities has been to obtain high-quality stellar spectroscopy
in external galaxies. For example, this gain in sensitivity is
evident for targets in M31, where new Gemini-GMOS spectroscopy
of luminous supergiants (Cordiner et al. 2010) makes light work of
observations which were challenging with 4\,m telescopes
(e.g. Bianchi et al. 1994; Herrero et al. 1994).  There are two
observational regimes in the current 8-10\,m era:
\begin{itemize}
\item{{\it Fundamental stellar astrophysics in nearby galaxies:}
High-quality spectroscopy for quantitative analysis of individual
massive stars in the Magellanic Clouds was only possible for selected
bright targets with 4\,m telescopes. With the advent of
multi-object instruments such as FLAMES on the VLT we have been able
to obtain large spectroscopic samples of the massive star populations
of the Clouds (and the Milky Way) to investigate the role of
environment on stellar properties and evolution (e.g. Evans et
al. 2005, 2006; Martayan et al. 2006, 2007).}

\item{{\it Stars as tracers of galaxy properties:}
Beyond 1\,Mpc, we have had our first glimpses of
spectra of individual stars in galaxies such as NGC\,3109 (Evans
et al. 2007) at 1.3\,Mpc, and NGC\,300 (Bresolin et al. 2002)
and NGC\,55 (Castro et al. 2008), both of which are spiral galaxies in the Sculptor `Group' at 
$\sim$1.9\,Mpc (Pietrzy\'{n}ski et al. 2006).  Such data can be used
to estimate the present-day abundances in these galaxies, 
to study radial abundance trends in the spirals (Urbaneja et al. 2005), and
to help refine our understanding of other diagnostics used in
interpretation of more distant systems (Bresolin et al. 2009).  The most
impressive observations in this context
are the VLT spectra of two supergiants in NGC\,3621 (Bresolin et
al. 2001), at a remarkable distance of 6.7\,Mpc.  }
\end{itemize}

The arrival of the ELTs will represent an even greater leap forward in
our capabilities than that when moving from 4 to 8\,m, opening-up 
an exciting range of options for future study, including:
\begin{itemize}
\item{{\it Physics \& evolution of massive stars in metal-poor
irregulars:} Quantitative analysis of blue
supergiants in galaxies such as IC\,1613, WLM and NGC\,3109 has found
oxygen abundances slightly below those found in the Small Magellanic
Cloud (Bresolin et al., 2006, 2007; Evans et al. 2007).  Determining
exact metallicities requires further work, but it is clear that these
systems provide an excellent opportunity to expand our studies of
massive stars in metal-poor regimes -- from the most extreme, luminous
phases to main-sequence dwarfs.}
\item{{\it Massive star populations of nearby spirals:}
The full luminosity range of massive stars will also be accessible in
local spirals.  This includes galaxies such as NGC\,55 and NGC\,300,
but the real prize here is the potential of studying the full
age-range of populations in both M31 and M33 -- from main-sequence
massive stars to the oldest evolved red giants.  There is already
significant deep imaging in these galaxies, and a new multi-cycle {\em
HST} Treasury Program to image one quadrant of M31 at
UV/optical/near-IR wavelengths (`A Panchromatic Hubble Andromeda
Survey; PI: J. Dalcanton) will provide rich sources of exciting
objects worthy of ELT spectroscopy for quantitative abundances and
radial velocities.}
\item{{\it The most luminous stars in distant galaxies:}
As we move beyond the Local Group, there is a rich assortment of
galaxy types and environments/groups.  Here we can seek to use our
understanding of massive stars to learn about the host galaxies. For
example, in the starburst galaxy M82, in ellipticals such as Cen~A
(NGC\,5128), NGC\,3379 and members of the Virgo Cluster, and in
interacting systems such as M51. Looking even further afield, we can
exploit the angular resolution of the ELTs to resolve the cluster
complexes in the Antennae into their sub-components to assess their
ages/populations.  Perhaps one of the most compelling targets for
further study in terms of its apparently {\em very} metal-poor
population is I~Zw\,18 (Heap et al., these proceedings), at a distance
18.2\,Mpc (Aloisi et al. 2007).  As noted by others in the literature,
I~Zw\,18 could provide important insights into stellar evolution in
conditions which are more in keeping with those in the very early
universe.}
\end{itemize}

\subsection{Synergies with other facilities}

The current generation of ground-based, optical-IR telescopes will
continue to deliver exciting new results over the coming decade,
particularly with the arrival of new instrumentation.  However, deep
imaging from the 8-10\,m telescopes and the {\em Hubble Space
Telescope (HST)} is already revealing targets which are beyond our
spectroscopic capabilities.  The need for follow-up will become
increasingly important, not to mention the potential of combining,
e.g, VLTI observations with ELT integral-field spectroscopy.

Such synergies will become even more crucial when looking ahead to
ALMA and the {\em James Webb Space Telescope (JWST)}, both of which
will start operations in the coming few years.  They will be unique at
their respective wavelengths, but supporting ground-based, optical-IR
observations will be critical, as exemplified by years of {\em HST}
operations.

Looking further ahead, the impact of other facilities such as the Large Synoptic
Survey Telescope (LSST), the Square Kilometre Array (SKA), and the
{\em International X-ray Observatory (IXO}; Rauw, these proceedings)
would all benefit hugely from the ability to obtain, e.g., spectroscopic follow-up or diffraction-limited
imaging with an ELT.

\subsection{Anticipating the unknown}
There are a wide range of scientific motivations for the ELTs, all
informed by our current research and contemporary understanding; it is
harder to plan for the unknown discoveries which await in the coming
decade. Harwit (1981) made the point that new discoveries are
generally achieved when a new part of parameter space is accessed for
the first time. An excellent example of this is provided by the first
sub-millimetre observations of distant galaxies (e.g. Smail, Ivison \&
Blain, 1997).  The ELTs will excel in the combination of collecting
power and angular resolution so, while we should design the
observatories and their instruments to provide the best possible
performance for the observations we can contemplate now, we should be
wary of focussing those capabilities too much, rendering us unable 
to investigate future discoveries that we can not even conceive of today.

\section{ELTs: Worldwide context}\label{elt}
Efforts toward building ELTs are becoming increasingly global, with
three projects now in the advanced stages of their
design, fund-raising and planning; the top-level details of each
observatory are summarised briefly below.  While these projects are collectively
referred to as `ELTs', note that there is a large range in the effective 
areas of their primary apertures.

\subsection{GMT: The Giant Magellan Telescope\footnote{http://www.gmto.org} }
The GMT employs seven monolithic 8\,m mirrors to form the primary aperture.
Six of these are off-axis, arranged around the central on-axis mirror (see left-hand panel of Fig.~\ref{gmt_tmt}).
The effective diameter in terms of potential angular resolution is 24.5\,m, with
an equivalent collecting area of a $\sim$22\,m filled-aperture primary.
At the time of writing, GMT includes ten partners. These are primarily
in the United States, but also include members in South Korea and
Australia.  The intended GMT site is at the Las Campanas Observatory in
Chile, already home to the Magellan, Du Pont and Swope telescopes.

\subsection{TMT: The Thirty Meter Telescope\footnote{http://www.tmt.org} }
\begin{center}
\begin{figure}[h]
\vspace{-0.2in}
\includegraphics{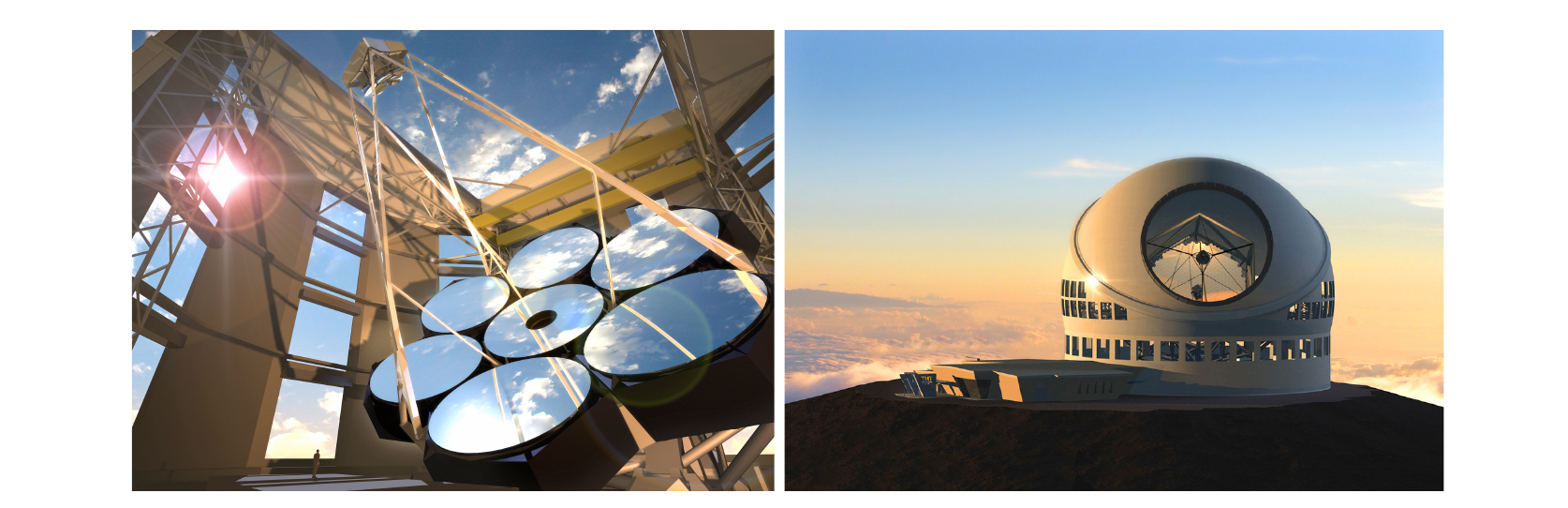}
\caption{{\it Left:} The GMT in its enclosure (image credit:
Giant Magellan Telescope -- GMTO Corporation); {\it Right:} Artist's interpretation 
of the TMT on Mauna Kea, Hawaii (credit: the Thirty Meter Telescope Project). \label{gmt_tmt}}
\end{figure}
\end{center}

The initial partners of TMT were Caltech, University of California and
Canada.  Over the past couple of years this partnership has expanded
to include Japan, China and, most recently, India.  The effective
diameter of the primary is 30\,m, comprised of hexagonal segments
which are just over 1.4\.m in diameter (across the corners). This
design builds on the considerable experience of the segmented
primaries of the Keck telescopes.  The TMT will be located on Mauna
Kea in Hawaii, with a novel `Calotte' dome to minimise wind shake (and
cost), as shown in the right-hand panel of Fig.~\ref{gmt_tmt}.  An
integral part of the project has been to minimise the environmental
impact of the observatory on the mountain, including an 
updated design of the dome and the offices (beyond the one shown in
Fig.~\ref{gmt_tmt}).

\subsection{E-ELT: The European Extremely Large Telescope\footnote{http://www.eso.org/sci/facilities/eelt} }
\begin{center}
\begin{figure}
\vspace{-0.2in}
\centering
\includegraphics[height=4.5cm]{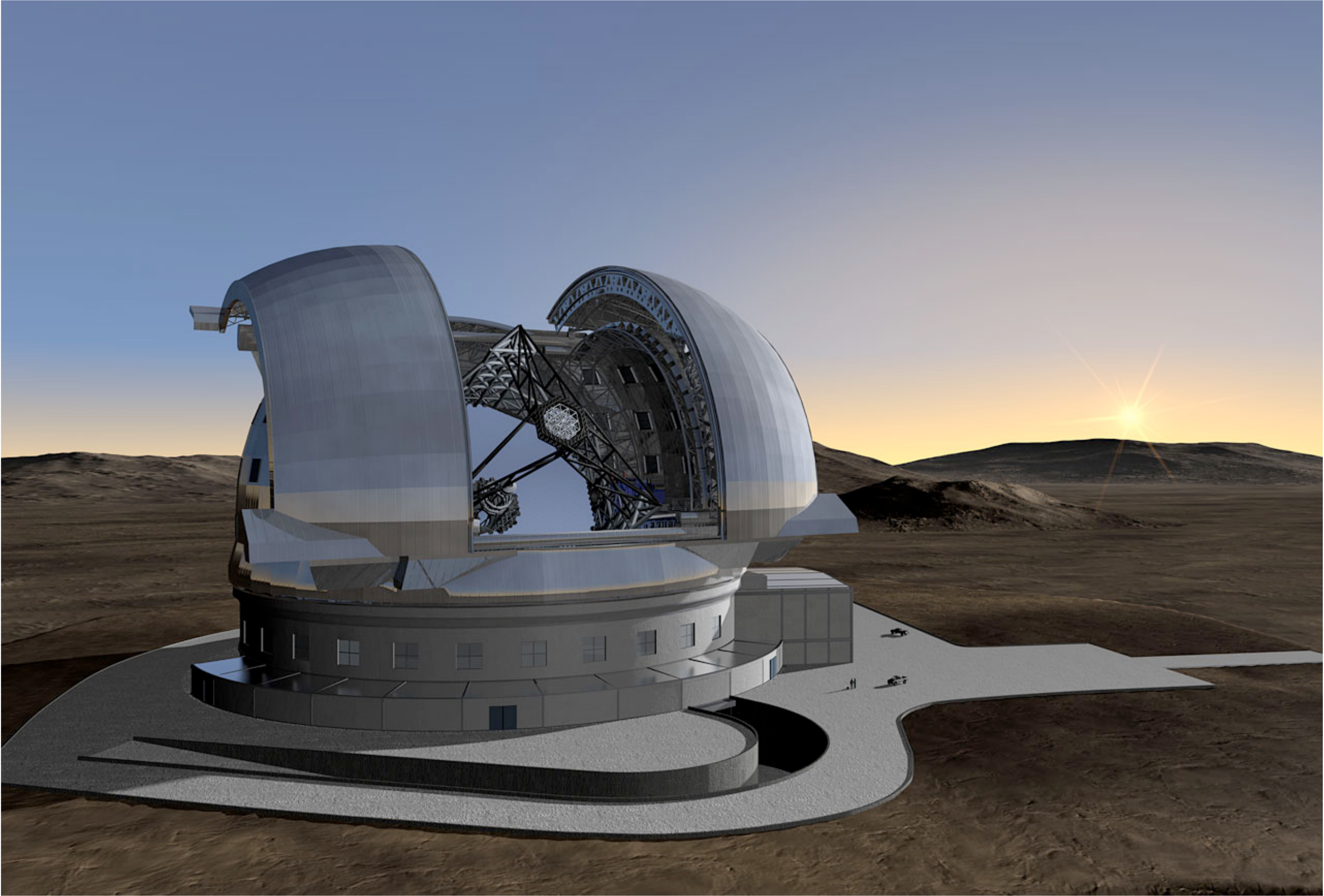}
\caption{Recent design of the E-ELT in its enclosure (Credit: ESO). \label{figeelt}}
\end{figure}
\end{center}

The E-ELT is under development by the European Southern Observatory
(ESO) on behalf of its partners, and features a primary with an
equivalent diameter of 42\,m (Fig.~\ref{figeelt}).  As with the TMT, the
primary is comprised of 1.4\,m hexagonal segments.
In addition to the primary, secondary and tertiary elements,
the telescope design features an adaptive fourth mirror and a fast
tip-tilt fifth mirror.  The E-ELT site was announced earlier in 2010 to be 
Cerro Armazones in northern Chile. This is approximately
20\,km from Paranal (the site of the VLT), meaning that some of the
infrastructure and operations costs can be shared.  Armazones is
at an altitude of 3060\,m, slightly higher than Paranal.

\subsection{Instrumentation plans}
Detailed overviews of the instrumentation studies completed or underway for
each of the three ELT projects were given at the recent SPIE meeting in 
San Diego (GMT: Jaffe et al. 2010; TMT: Simard et al. 2010; E-ELT: Ramsay et al. 2010). 
A vast range of parameter space is covered by these studies.  This is, in 
part, to be able to evaluate the relative merits of different capabilities
toward each science case, but also to explore the technology 
readiness/requirements of key components for the future. 

The instrument studies undertaken to date are given in Tables~\ref{gmtins},
\ref{tmtins}, and \ref{eeltins}.  Part of the motivation for such a comprehensive list is 
not to dazzle or overwhelm with acronyms and abbreviations, but to
illustrate the significant effort that has already gone into these
projects from the instrumentation part of the community, in close
collaboration with the astronomers who have developed the science cases,
undertaken simulations etc.

The first round of studies for TMT were completed in 2008, with
three instruments subsequently selected for `early light' operations,
as indicated in Table~\ref{tmtins}. A similar down-select will form part
of the E-ELT construction proposal, to be released in late 2010.  The
GMT studies will end in July 2011, with a down-select following thereafter.

\begin{table}[h!]
\caption{GMT conceptual design studies (Jaffe et al. 2010).}\label{gmtins}
\footnotesize
\vspace{-0.15in}
\begin{center} 
\begin{tabular}{| p{2cm} | p{11cm} |}
\hline 
Instrument & Brief description\\
\hline
G-CLEF & High resolving power, high stability, optical spectrograph \\
GMACS & Multi-slit, seeing-limited, optical spectrograph \\
GMTIFS & AO-corrected, integral field, near-IR spectrogragh ({\it \`{a} la} GEMINI-NIFS) \\
GMTNIRS & AO-fed, high resolving power, near-IR spectrograph \\
NIRMOS & Multi-slit, near-IR spectrograph \\
TIGER & Mid-IR imager and low-resolution spectrograph \\
\hline 
\end{tabular} 
\end{center} 

\caption{TMT capabilities for first decade (Simard et al. 2010).}\label{tmtins}
\footnotesize
\vspace{-0.15in}
\begin{center} 
\begin{tabular}{| p{2cm} | p{11cm} |}
\hline 
Instrument & Brief description\\
\hline
IRIS & AO-fed, near-IR, integral field unit (IFU) and imager \\
IRMS & AO-fed, multi-slit, near-IR spectrograph (clone of Keck-MOSFIRE) \\
WFOS & Seeing-limited, multi-object, optical spectrograph \\
\hline
HROS & High-resolving power, seeing-limited, optical spectrograph \\
IRMOS & Multi-IFU, AO-corrected, near-IR spectrograph\\
MIRES & AO-fed, mid-IR, echelle spectrograph \\ 
NIRES & AO-fed, near-IR, echelle spectrograph \\
PFI & High contrast, near-IR imager \\
WIRC & `Wide field', AO-corrected imager \\
\hline
\end{tabular} 
\footnotesize{~\\ ~\\}The first three instruments (IRIS, IRMS and WFOS) are those planned for `early light'.
\end{center} 
\end{table} 

\begin{table}
\caption{E-ELT Phase A studies (Ramsay et al. 2010).}\label{eeltins}
\footnotesize
\vspace{-0.15in}
\begin{center} 
\begin{tabular}{| p{2cm} | p{11cm} |}
\hline 
Instrument & Brief description\\
\hline
CODEX & High resolving power, high stability, optical spectrograph \\
EAGLE & Multi-IFU, AO-corrected, near-IR spectrograph \\
EPICS & High contrast, near-IR imager/spectro-polarimeter \\
HARMONI & AO-fed, near-IR, IFU \\
METIS & AO-fed, mid-IR imager and spectrograph \\
MICADO & Near-IR, diffraction-limited imager \\
OPTIMOS & Seeing-limited/ground-layer AO, high-multiplex spectrograph \\
SIMPLE & AO-fed, near-IR, high resolving power spectrograph \\
\hline
\end{tabular} 
\end{center} 
\end{table} 

\newpage
\vspace{-0.2in}
\section{Illustrative performances}\label{performances}

To illustrate the spectroscopic potential of ELTs,
I refer to simulations from the EAGLE Phase A study
for the E-ELT (Cuby et al. 2010).  EAGLE is a conceptual design for an
AO-corrected, near-IR spectrograph with multiple integral field units
(IFUs).  A key element of its science case is spectroscopy of resolved
stellar populations beyond the Local Group, using evolved
red giant stars to trace the star-formation histories of their host
galaxies.  Tools have been developed to simulate
EAGLE observations (Puech et al. 2008, 2010), which employ a set of
simulated, AO-corrected, point-spread functions (PSFs) for example configurations
of natural guide stars (Rousset et al. 2010).

Simulated EAGLE performances for spectroscopy of the calcium triplet
(centered at 0.86\,$\mu$m) were given by Evans et al. (2010).  The
continuum signal-to-noise (S/N, per pixel) resulting from some of these
simulations is summarised in Table~\ref{cat} for two configurations
of guide stars; other relevant parameters were a
spectral resolving power, $R$, of 10000, a total exposure time of
10\,hrs (20$\times$1800s) and a seeing of 0\farcs65.  From a stacked
10\,hr exposure at $I$\,$=$\,24.5 (in the Vega System), a continuum
S/N\,$\ge$\,10 is recovered, some four magnitudes deeper than
FLAMES-GIRAFFE on the VLT using the LR08 setting (which is also at a
lower resolving power of $R$\,$=$\,6500), with the same exposure time.

\vspace{-0.15in}
\begin{table}[h!]
\caption{Continuum signal-to-noise (S/N, per pixel) obtained for simulated EAGLE observations of
the calcium triplet (Evans et al. 2010), with $R$\,$=$\,10000, $t_{\rm
exp}$\,$=$\,10\,hrs, and two configurations of natural guide stars (NGS).}\label{cat}
\footnotesize
\vspace{-0.15in}
\begin{center} 
\begin{tabular}{| c | c | c |}
\hline 
& S/N & S/N \\
$I_{\rm VEGA}$ & [NGS `Good'] & [NGS `Poor'] \\
\hline
22.5 & 56 & 48 \\
23.5 & 28 & 24 \\
24.5 & 13 & 10 \\
\hline
\end{tabular} 
\end{center} 
\end{table} 

\vspace{-0.2in}
These simulations were originally to quantify the performances for red
giant stars, but we can also consider their implications in the
context of massive stars, where we typically require a S/N\,$\ge$\,50.
The reach of $I$-band spectroscopy of different populations is
summarised in Table~\ref{sndist}.  I know the $I$-band is not exactly
replete with useful diagnostics in the spectra of massive stars (!),
but these performances are merely to give a feel for the distances to
which one can contemplate ELT observations.  The bottom-line is that
the ELTs will provide spectroscopy of individual stars beyond the
Local Group, in the same manner that we have begun to take for granted
in nearby galaxies.

\begin{table}[h]
\caption{Distance moduli (DM) and distances ($d$) to which $I$-band spectroscopy
could be obtained with EAGLE (at $R$\,$=$\,10000 and given S/N),
assuming the performances from Table~\ref{cat}.}\label{sndist}
\footnotesize
\vspace{-0.15in}
\begin{center}
\begin{tabular}{| l | c | ccc |}
\hline
Star & S/N & M$_I$ & DM & $d$ \\
\hline
Tip of red giant branch & $\ge$\,10 & $-$4$\phantom{.0}$ & 28.5 & $\phantom{.0}$5~Mpc \\
BA-type supergiants & $\ge$\,50 & $-$7$\phantom{.0}$ & 29.5 & $\phantom{.0}$8~Mpc \\
O-type dwarf & $\ge$\,50 & $-$4.5 & 27.0 & 2.5~Mpc \\
\hline
\end{tabular}
\end{center}
\end{table}

\subsection{Future diagnostics}
\begin{center}
\begin{figure}[h]
\vspace{-0.25in}
\centering
\includegraphics[height=9cm,angle=-90]{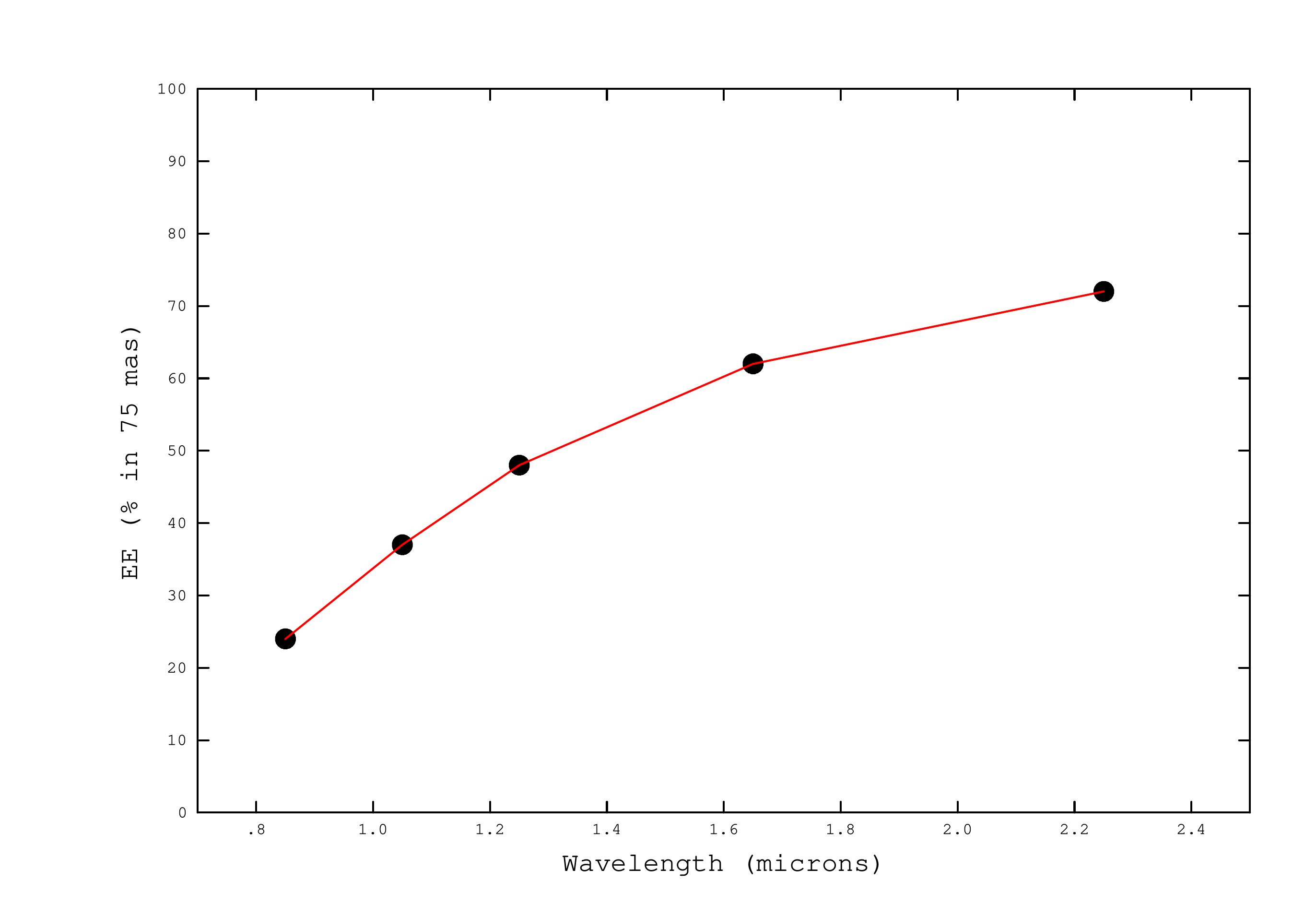}
\caption{The wavelength dependence of the effective AO correction is shown by 
the encircled energy (EE) delivered in 75\,milliarcseconds by the
simulated EAGLE PSFs (Rousset et al. 2010).\label{aoperf}}
\end{figure}
\end{center}

An important additional factor in performance calculations for ELTs is
the level of AO correction.  Seeing-limited spectroscopy in the
optical (covering the diagnostic lines with which we have considerable
experience) could yield spectra of supergiants in the outer halos of
distant galaxies, but to probe into the densest regions we will
require the best possible contrast from AO, which will be limited to
relatively small fields.
Moreover, for stellar spectroscopy with the ELTs there will be a fine
balance in sensitivity between the improved image quality from
adaptive optics as one goes to longer wavelengths (where the wavefront
errors become less significant compared to the observational
wavelengths) versus the increased background.  As an example
of the wavelength dependence of the AO correction, Fig.~\ref{aoperf}
shows results for the encircled energy within a 75\,milliarcsecond
aperture from the simulated EAGLE PSFs (Rousset et al. 2010).

\vspace{-0.35in}
Martins (these proceedings) summarised the broad range of diagnostics
available to us in our efforts to determine the physical parameters of
massive stars.  For example, in regions of high extinction such as the
Galactic Centre, $K$-band spectroscopy provides the means to study
stellar winds (Martins et al. 2007).  Work is already underway to
improve our understanding of near-IR diagnostics compared to those in
the optical as part of the planning toward ELTs (e.g. Przybilla et
al. 2009; Nieva et al. 2009), not to mention compilation of a
high-resolution spectral library in the near-IR (Lebzelter et
al. 2010; Ramsay et al., these proceedings).

\begin{center}
\begin{figure}[h]
\vspace{-0.2in}
\centering
\includegraphics[height=8cm]{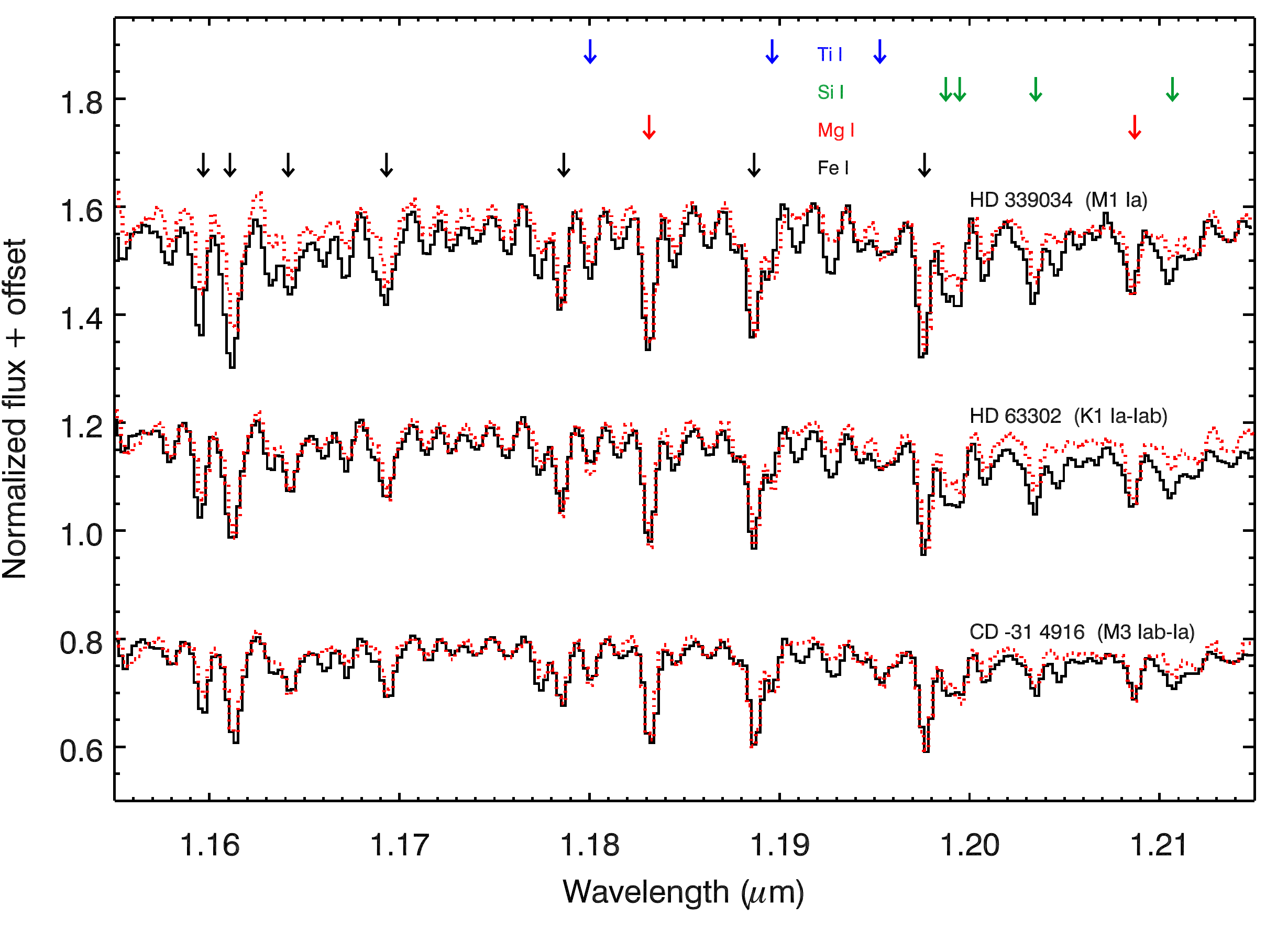}
\caption{Model atmosphere fits (red lines) to provide metallicities/stellar 
abundances from $J$-band spectra of Galactic red supergiants (Davies, Kudritzki
\& Figer, 2010).\label{bd3}}
\end{figure}
\end{center}

A more detailed example of exploring alternative diagnostics is the
recent study by Davies, Kudritzki \& Figer (2010) in which they employ
low-resolution ($R$\,$\sim$\,2-3000) $J$-band spectroscopy to determine
metallic abundances for red supergiants -- this is a relatively
unexplored wavelength domain, first considered for this type of
abundance work by Origilia et al. (2004).  From comparisons with contemporary
model atmospheres, the method provides good estimates of stellar
metallicities (with a dispersion of $\pm$0.14 dex) for library spectra
in the Solar neighbourhood; example fits are shown in Fig.~\ref{bd3}.
We are now exploring the potential of this method for metal-poor templates
in simulated ELT observations, using the tools mentioned earlier that were 
developed for EAGLE.  By their nature, red supergiants are cool
(i.e. significant $J$-band flux in their spectral energy distributions) and
very luminous - when combined with the AO capabilities of the ELTs, this
method could provide a probe of stellar abundances out to distances of tens
of Mpc.

\section{MAD: An AO pathfinder for ELTs}\label{mad}

Correction for the effects of atmospheric turbulence with AO is a
critical ingredient of the plans for ELTs and their instruments.  In
particular, there is a strong desire for good and {\em uniform}
correction over larger fields-of-view than delivered by, e.g.,
VLT-NACO (Rousset et al. 2003). A key technical component within ESO's
plan toward the E-ELT was an on-sky demonstration of multi-conjugate
adaptive optics (MCAO). To this end, the Multi-conjugate Adaptive
optics Demonstrator (MAD) was developed (Marchetti et al. 2007)
with commissioning at the VLT in early 2007.

MAD employs three wavefront sensors to observe three natural guide
stars across a 2$'$ circular field, thereby allowing tomography of the
atmospheric turbulence.  The turbulence is then corrected using two
deformable mirrors, one conjugated to the ground-layer (i.e. 0\,km),
the second conjugated to 8.5\,km above the telescope.  The MAD near-IR
camera critically samples the diffraction-limited PSF at 2.2$\mu$m,
giving a pixel scale of 0\farcs028/pixel. With a 2k$\times$2k Hawaii-2
array, this gives a total field-of-view of 57$''$\,$\times$\,57$''$.

\begin{figure}[h]
\centering
\includegraphics[width=10cm]{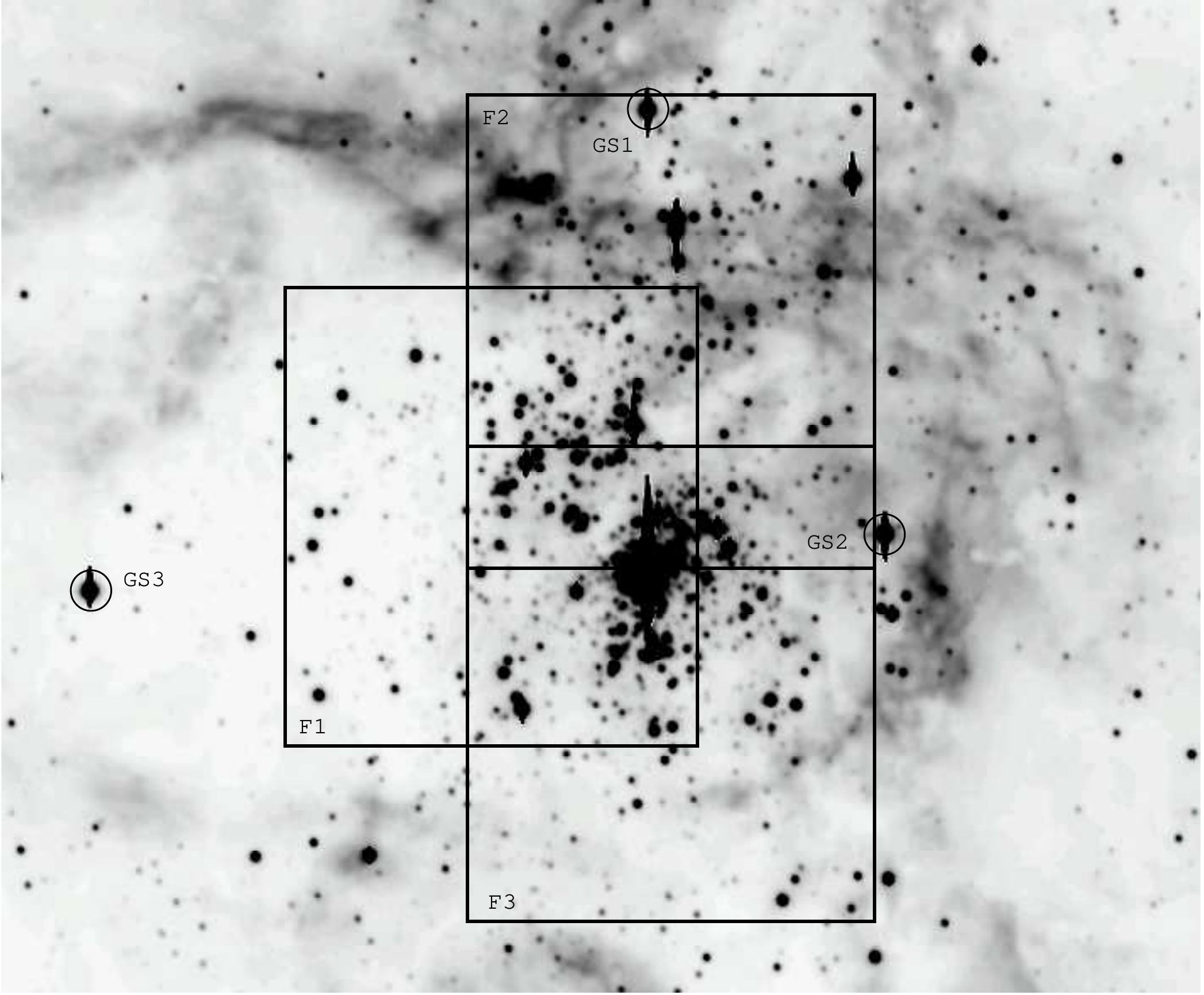}
\caption{The three MAD pointings overlaid on a $V$-band image of the
central part of 30~Dor.  North is at the top, east toward the left.  The three reference
stars used for the AO correction are shown (`GS\#').\label{fig1}}
\end{figure}

The commissioning performances were sufficiently compelling that ESO
issued a call for Science Demonstration (SD) observations, to which
the community responded enthusiastically.  As an illustration of the
SD programmes I refer to $H$- and $K_{\rm s}$-band MAD observations of
R136 (Fig.~\ref{fig1}; Campbell et al. 2010), the dense cluster at the
core of 30~Doradus in the Large Magellanic Cloud.  Also of direct
relevance to the topic of this meeting are the MAD observations of
Trumpler~14 (Sana et al. 2010).

\begin{figure}[h]
\centering
\includegraphics[width=10cm]{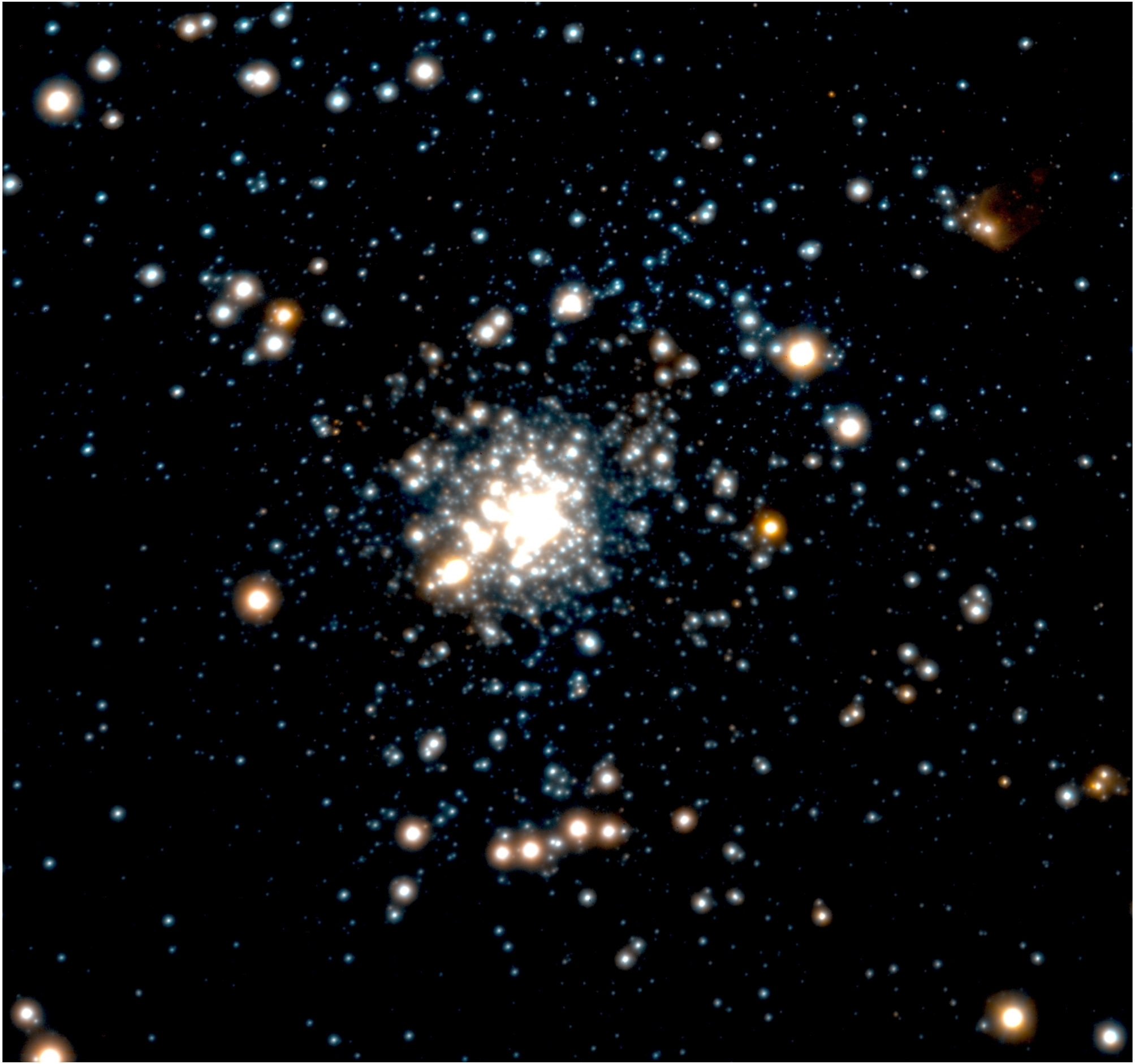}
\caption{Combined mosaic of the central 40$''$\,$\times$\,40$''$ MAD images (blue: $H$; red: $K_{\rm s}$;
green: interpolated). \label{fig2}}
\end{figure}

The AO correction is such that the mean full-width half-maximum of the PSFs
is $\sim$0\farcs10 in two of the
three MAD pointings (the best placed with respect to the guide stars).  This
provides near-comparable angular resolution to optical imaging with
the {\em HST} (e.g., Hunter et al. 1995). Due to the size of the primaries, 
the angular resolution from MAD is finer than that from the {\em HST} at the
same wavelength, although the {\em HST} has a large advantage in terms
of sensitivity due to the reduced sky background (cf. the results of
Andersen et al. 2009).  Further details regarding calibration and
performance are given by Campbell et al. (2010); a combined image
of the central region is shown in Fig.~\ref{fig2}.

A nice test of the ground-based methods with MAD, compared to the {\em
HST} results from Andersen et al., is provided by the agreement in the
derived slopes of power-law fits to the luminosity profile of R136.
The MAD data also have the advantage of going out to larger cluster
radii than those from Andersen et al., allowing re-investigation of a
`bump' seen in the luminosity profile of R136 from optical {\em HST}
observations (Mackey \& Gilmore, 2003).  This `excess light' at large
radii has been suggested in the past as perhaps related to the
signatures of rapid gas expulsion from young clusters (e.g. Goodwin \&
Bastian, 2006).  In contrast, the MAD data do not reveal an obvious
break in the luminosity profile, with Campbell et al. (2010)
suggesting that cluster asymmetries are the dominant source.  In
combination with AO-corrected IFU spectroscopy (Schnurr et al. 2009),
the MAD data have since been used to argue that the central stars of
R136 have initial masses in excess of 150\,M$_\odot$ (Crowther et
al. 2010; see also Schnurr et al., these proceedings).

The instrumental and scientific experiences of MAD augur well for
the AO plans for ELTs.  Pathfinders for other AO modes such as
Multi-Object Adaptive Optics are underway, for example the CANARY
project (Myers et al. 2008; Morris et al. 2010) and plans for RAVEN
(Conan et al. 2010).  In the shorter term, the Gemini MCAO System
(GeMS) is now undergoing its final tests and integration (Neichel et
al. 2010), and will offer a unique capability of MCAO with laser guide
stars (removing some of the constraints on natural guide stars), for
near-IR imaging and multi-object spectroscopy.

\section{Summary}
The ELTs offer huge potential for studies of massive stars -- we will
be able to address fundamental questions regarding their formation and
evolution (with a particular focus on environmental effects), while
also using individual stars as tracers of the stellar populations in
galaxies well beyond the Local Group.

By way of further motivation, consider the ground-breaking study by
Swinbank et al. (2010) in which they have used multi-wavelength
observations to study a gravitationally-lensed galaxy at a redshift of
$z$\,$=$\,2.3. By virtue of the lens, individual star-forming regions
are resolved in sub-millimetre imaging, each $\sim$100\,pc in scale --
i.e. direct observation of intense regions of star formation with
spatial extents comparable to that of 30~Doradus, at a time when the
universe was significantly younger. Such an observation is only
possible at present due to the magnification of the lens but, with the
combined power of ALMA and the ELTs in the future, we can expect
comparable observations in unlensed systems. One of the
challenges ahead is to improve our models of stellar evolution to the
point at which we are confident that we can interpret integrated-light
spectroscopy of such distant systems accurately, thereby exploiting our
understanding of massive stars to obtain new insights into the
processes at work during the critical epoch of galaxy evolution.

%
%
\section*{Acknowledgements}
Thanks to Michael Campbell for his MAD mosaic and to Ben Davies for copies of his figures.

%
%
\footnotesize
\beginrefer
\refer Aloisi, A., et al. 2007, ApJ, 667, L151

\refer Andersen, M., et al. 2009, ApJ, 707, 1347

\refer Bianchi, L., et al. 1994, A\&A, 292, 213

\refer Bresolin, F., et al. 2001, ApJ, 548, L159

\refer Bresolin, F., et al. 2002, ApJ, 567, 277

\refer Bresolin, F., et al. 2006, ApJ, 648, 1007

\refer Bresolin, F., et al. 2007, ApJ, 671, 2028

\refer Bresolin, F., et al. 2009, ApJ, 700, 309

\refer Campbell, M. A., et al. 2010, MNRAS, 405, 421

\refer Conan, R., et al. 2010, SPIE, 7736, 26

\refer Cordiner, M., et al. 2010, ApJ, in press

\refer Crowther, P. A., et al. 2010, MNRAS, arXiv:1007.3284

\refer Cuby, J.-G., et al. 2010, SPIE, 7735, 80

\refer Davies, B., Kudritzki, R.-P. \& Figer, D. F., 2010, MNRAS, 407, 1203

\refer Davies, B., et al. 2010, MNRAS, 402, 1504

\refer de Wit, W. J., et al. 2007, ApJ, 671, L169

\refer de Wit, W. J., et al. 2010, A\&A, 515, A45

\refer Evans, C. J., et al. 2005, A\&A, 456, 623

\refer Evans, C. J., et al. 2006, A\&A, 437, 467

\refer Evans, C. J., et al. 2007, ApJ, 659, 1198

\refer Evans, C. J., et al. 2010, in Cl\'{e}net, Conan, Fusco \& Rousset, eds, {\it
Adaptive Optics for Extremely Large Telescopes}, EDP Sciences, 1004, arXiv:0909.1748

\refer Gies, D. R., 2008, in Beuther, Linz \& Henning, eds, {\it Massive
Star Formation: Observations Confront Theory}, ASP Conference Series, 387, p93

\refer Goodwin, S. P. \& Bastian, N., 2006, MNRAS, 373, 752

\refer Harwit, M. 1981, {\it Cosmic Discovery}, Basic Books, New York

\refer Herrero, A., et al. 1994, A\&A, 287, 885

\refer Hunter, D. A., et al. 1995, ApJ, 448, 179

\refer Jaffe, D., et al. 2010, SPIE, 7735, 72

\refer Kraus, S., et al. 2010, Nature, 466, 339

\refer Lebzelter, T., et al. 2010, Msngr, 139, 33

\refer Mackey, A. D. \& Gilmore, G. F., 2003, MNRAS, 338, 85

\refer Marchetti, E., et al. 2007, Msngr, 129, 8

\refer Martayan, C., et al. 2006, A\&A, 452, 273

\refer Martayan, C., et al. 2007, A\&A, 462, 683

\refer Martins, F., et al. 2007, A\&A, 468, 233

\refer Mason, B. D. et al. 2009, AJ, 137, 3358

\refer Morris, T., et al. 2010, in Cl\'{e}net, Conan, Fusco \& Rousset, eds, {\it
Adaptive Optics for Extremely Large Telescopes}, EDP Sciences, 8003

\refer Myers, R. M., et al. 2008, SPIE, 7015, 6

\refer Nieva, M. F., et al. 2009, in Moorwood, ed., {\it Science with the VLT in the ELT era}, 
Springer, Netherlands, p499

\refer Neichel, B., et al. 2010, SPIE, 7736, 4

\refer Origlia, L., et al. 2004, ApJ, 606, 862

\refer Pietrzy\'{n}ski, G., et al. 2006, AJ, 132, 2556

\refer Przybilla, N., et al. 2009, in Moorwood, ed., {\it Science with the VLT in the ELT era}, 
Springer, Netherlands, p55

\refer Puech, M., et al. 2008, MNRAS, 390, 1089

\refer Puech, M., et al., 2010, SPIE, 7735, 183

\refer Ramsay, S., et al. 2010, SPIE, 7735, 71

\refer Rousset, G., et al. 2003, SPIE, 4839, 140

\refer Rousset, G., et al. 2010, in Cl\'{e}net, Conan, Fusco \& Rousset, eds, {\it
Adaptive Optics for Extremely Large Telescopes}, EDP Sciences, 2008, arXiv:1002.2077

\refer Sana, H., et al. 2010, A\&A, 515, 26

\refer Sana, H. \& Evans, C. J., 2010, to appear in Neiner, Wade, Meynet \& Peters, eds, Proc. IAUS272:
{\it Active OB Stars: Structure, Evolution, Mass loss \& Critical
Limits}, Cambridge University Press

\refer Schnurr, O., et al. 2009, MNRAS, 397, 2049

\refer Simard, L., et al. 2010, SPIE, 7735, 70

\refer Smail, I., Ivison, R. J. \& Blain, A. W., 1997, ApJ, 490, L5

\refer Swinbank, A. M., et al. 2010, Nature, 464, 733

\refer Urbaneja, M. A., et al. 2005, ApJ, 635, 311

\refer Zinnecker, H., 2006, in Whitelock, Dennefeld \& Leibundgut, eds, Proc. IAUS232: {\it The
Scientific Requirements for Extremely Large Telescopes}, Cambridge University Press, p324

\endrefer           
\end{document}